\documentclass[aps,prl,twocolumn]{revtex4-1}
\usepackage{graphicx}
\usepackage{amsmath,verbatim,amssymb,stmaryrd}
\usepackage{hyperref}
\usepackage{xcolor}

\begin{document}
	

\title{Polarized light from the transportation of a matter-antimatter beam in a plasma}
\author{Ujjwal Sinha}
\author{Christoph H. Keitel}
\author{Naveen Kumar}\email[Corresponding author: ]{naveen.kumar@mpi-hd.mpg.de}\affiliation{Max-Planck-Institut f\"ur Kernphysik, Saupfercheckweg 1, D-69117 Heidelberg, Germany}
	
\date{\today}

\begin{abstract}
	A relativistic electron-positron beam propagating through a magnetized electron-ion plasma is shown to generate both circularly and linearly polarized synchrotron radiation. The degrees of circular and linear polarizations depend both on the density ratio of pair beam to background plasma and initial magnetization, and a  maximum degree of circular polarization $\langle P_\textrm{circ}\rangle \approx 18\%$ is found to occur for a tenuous pair beam. We demonstrate that the generation of circularly polarized radiation is intrinsically linked to asymmetric energy dissipation of the pair beam during the filamentation instability dynamics in the electron-ion plasma. These results can help in understanding the recent observations of circularly polarized radiation from gamma-ray-bursts.
\end{abstract}
\maketitle

Radiation from extreme astrophysical scenarios as \emph{e.g}  gamma-ray bursts (GRBs) occurring in supernova explosions, pulsar wind nebulae and active galactic nuclei is of immense interest as it holds vital clues about stellar dynamics and acceleration of cosmic rays~\cite{mundell2013,wiersema2014,band1993batse,hededal2005}. In particular, the radiation observed in the afterglow of GRBs, which are the most luminous events in the Universe, is central to understanding the role played by various nonlinear processes occurring in the astrophysical plasmas providing extremely efficient energy conversion of the matter into radiation and vice-versa.

GRBs are one of the unsolved problems in astrophysics, and they are capable of releasing enormous amount of energy ($E_k \sim 10^{51}$ erg) in a small volume of radius $\sim 10^{3}$ km ~\cite{Piran:1997aa}. Thus, one can expect a copious amount of pair-plasma ($e^{-},e^{+}$) being generated, which is termed as a fireball that propagates in the plasma made of baryonic matter~\cite{piran2005}. This can give rise to the onset of plasma instabilities, notably the Weibel instability (WI)~\cite{weibel1959} that can generate a strong magnetic field, converting the kinetic energy of the streaming particle into electromagnetic fields~\cite{Silva2003, muggli2013}. Both the pair-plasma and the relativistic ejecta of supernova explosion are capable of launching internal and external (with the interstellar medium) collisionless shocks, respectively. These shocks can accelerate the baryonic matter to high-energy and can be responsible for the ultra-high energy cosmic rays. The presence of a strong magnetic field can cause ultra-relativistic  particles to emit synchrotron radiation. Indeed, the radiation signatures from GRBs indicate the presence of strong equipartition magnetic field and a power law distribution of the cosmic rays~\cite{band1993batse, Begelman1984, Blandford:1977aa}. Moreover, the radiation is deemed to be linearly polarized~\cite{Rybicki:1985aa,Jackson:1998aa,Medvedev:1999aa}. However, recent observation of the circular polarized radiation~\cite{wiersema2014} in the afterglow of the GRBs calls for  studying the polarization properties of the radiation from GRBs and the role of plasma composition and dynamics  on the synchrotron radiation emission~\cite{mundell2013, [ {U. Sinha, J. Martins, J. Vieira,  K. M. Schoeffler, R. A. Fonseca, and L. Silva, (to be submitted)}. See also; ] [{. The approach of tracking particle trajectories from PIC simulation to compute the polarized radiation was first employed here, while the physical configuration and underlying physical processes are different in both manuscripts.} ] Sinha:2015aa, Sagiv:2004aa}. 

Parallel to the observations of astrophysical processes from ground based and airborne telescopes, there has also been a growing interest in utilising powerful and highly energetic laser systems~\cite{Boehly:1995aa,Lobet:2017aa, lobet2015,eli,cilex} to study collisionless shocks, magnetic reconnection and plasma instability dynamics in a controlled laboratory experiment~\cite{huntington2015,Sarri:2015aa,sarri2013,lobet2015,fox2013,warwick2017}. It is a rapidly evolving area of research known as laboratory astrophysics, which invokes the scale invariance principle to felicitate comparison between the laboratory experiments and the astrophysical scenarios.
Recently, a breakthrough experiment was conducted in which a dense pair plasma beam of density $n_b\sim 10^{16}$cm$^{-3}$ and an average Lorentz factor, $\gamma_\textrm{av}\approx 15$, was produced in a laboratory~\cite{Sarri:2015aa}. The generation of neutral high-density pair plasma is very encouraging as it can attempt to give insight into the transport of matter-antimatter beam in a plasma, mimicking the conditions of the fireball beam propagation in GRBs.

Motivated by these developments, in this Letter, we study the polarized radiation from propagation of a pair-plasma in a magnetized electron-ion plasma by 3D particle-in-cell (PIC) simulations. Here, the case of magnetized scenario is important as the polarization of GRBs indicate the presence of an ordered magnetic field ~\cite{mundell2013, wiersema2014}. The motives of our study are twofolds: First, it is analogous to the finite fireball beam propagation in a magnetized background plasma of baryonic matter and it can shed light on the observation of circular polarization in GRBs. Second, it compliments the ongoing experimental efforts in laboratory astrophysics, where the proton radiography technique is  heavily used to map the field strength and particle densities~\cite{fox2013, huntington2015}. Polarized radiation from plasma particle can also serve an important role in deciphering the dynamics of plasma instabilities, complimenting the proton radiography technique.

Full 3D PIC simulations on the propagation of a cold relativistic pair beam ($e_b^-, e_b^+$) in a preformed uniform electron-proton plasma are carried out using the electromagnetic PIC code SMILEI~\cite{derouillat2018smilei}. A simulation window of size $[16\times 32\times 32] ({c/\omega_\textrm{p}})^3$ and a resolution of $[256\times 256\times 256]$ cells moving with the speed of light along the $x$-direction with absorbing boundary conditions for particles and fields in the transverse direction was used. Here $\omega_\textrm{p}=(4\pi n_\textrm{0}e^2/m_\textrm{e})^{1/2}$ is the electron plasma frequency of the ambient electron-proton plasma, $n_\textrm{0}$ the electron density, $e$ and $m_\textrm{e}$ are the electronic charge and mass respectively, and $c$ is the velocity of the light in vacuum. The pair beam with a Gaussian density profile in each direction was initialized at the center of the simulation window with FWHM $[2\times 8.5\times 8.5 ]({c/\omega_\textrm{p}})^3$. We used the simulation timestep of $\Delta t = 0.0295 \omega_\textrm{p}^{-1}$ and $[2\times 2\times 2]$ particles per cell per species~\footnote{Higher number, \emph{e.g.} 100 particles per cell  produced similar results.}. The ambient plasma had  realistic proton to electron mass ratio, $m_\textrm{p}/m_\textrm{e}=1836$ with temperature $\sim$5 keV. The ambient plasma was magnetized with a uniform magnetic field $\textbf{B}_\textrm{0}=B_\textrm{0}\hat{\textbf{e}}_x$, of strength $B_\textrm{0}=[0.1, 0.2, 0.3, 0.4, 0.5, 0.75, 1.0]$ (normalized to $m_\textrm{e}\omega_\textrm{p}c/e$). For plasma density of $n_\textrm{0}=10^{16}\,\textrm{cm}^{-3}$, it corresponds to $B_0 \sim (3-30)$ T. These parameters are readily available for current laboratory astrophysics experiments and can be scaled up for comparison with astrophysical scenarios~\cite{Ryutov:2018aa}. \textcolor{black}{The choice of parallel magnetic field can correspond to either the magnetic field of central engine of the GRB, interstellar magnetic field compressed at the shock front or field generated due to other dynamical processes \cite{piran2005,Sagiv:2004aa,Medvedev:1999aa}.}

\begin{figure}
	\centering
	\noindent\includegraphics[width=0.49\textwidth, height=0.45\textwidth]{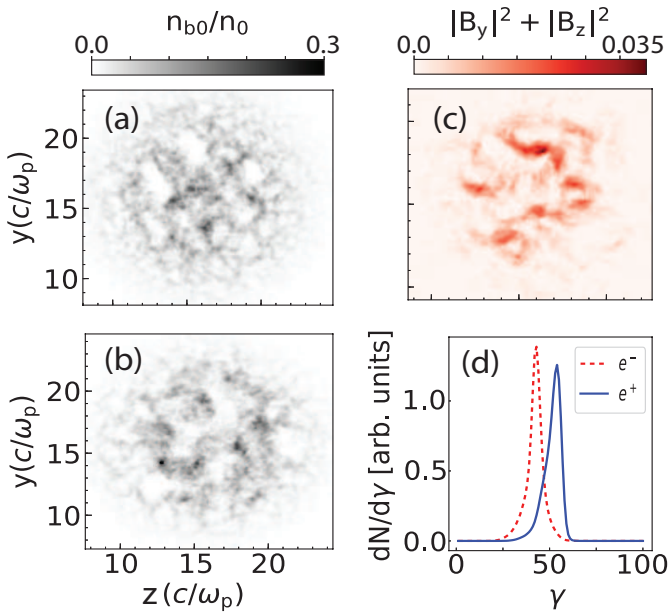}
	\caption{(a) and (b) $e_b^-$ and $e_b^+$ densities in the $yz$ plane respectively at $x=8 c/\omega_\textrm{p}$. (c) Spatial distribution of transverse magnetic field energy in the $yz$ plane at $x=8c/\omega_\textrm{p}$.(d) beam $e_b^-$(red) and the beam $e_b^-$(blue) energy distribution showing the deceleration of $e_b^-$ and acceleration of $e_b^+$.}\label{filaments}
\end{figure}

\begin{figure}
	\centering
	\noindent\includegraphics[width=0.45\textwidth, height=0.16\textwidth]{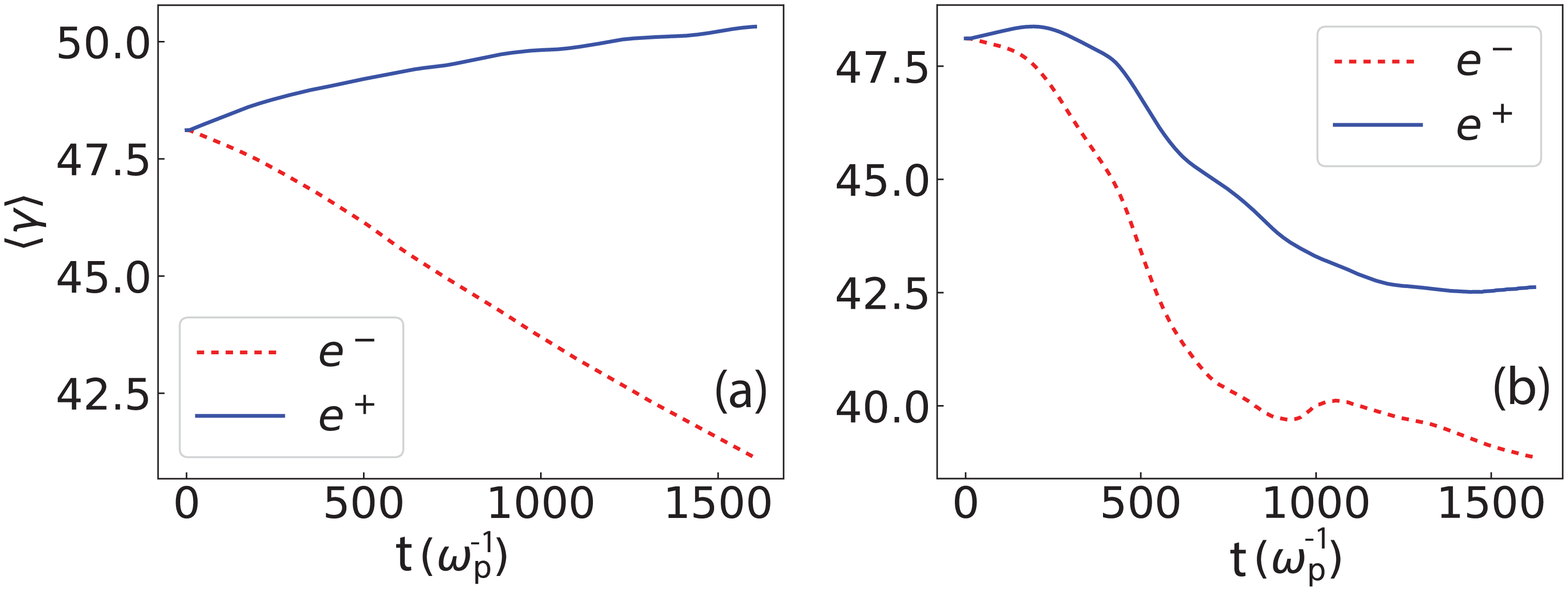}
	\caption{Energy dissipation of each species of the pair beam at two different beam to plasma density ratios (a) $n_{b0}/n_0=0.1$ and ($n_{b0}/n_0=1$). Other parameters are same as in Fig.\ref{filaments}.}
\label{enerdissip}
\end{figure}

When a pair beam ($e_b^-,e_b^+$) propagates in a plasma, the beam electrons ($e_b^{-}$) repel and the beam positrons ($e_b^{-}$) attract the background plasma electrons respectively, violating the quasi-neutrality of the plasma. This causes the generation of an inductive electric field that, in turn, generates a return plasma current to neutralize the effect of external beam. Since the pair-beam has transverse size larger than the plasma skin-depth $d_s=c/\omega_p$, the return plasma current can penetrate, and flow inside the pair-beam. This configuration of the opposite beam currents is unstable in a plasma and it can split the pair-beam into smaller filaments of size $\sim d_s$ due to the WI or current filamentation instability (WI/CFI), leading to the generation of a strong transverse magnetic field at the expense of the pair-beam energy.  Since in the case of an ambient electron-proton plasma, only the background electrons participate in the current neutralization as $m_\textrm{p}\gg m_\textrm{e}$, the $e_b^+$ and $e_b^-$  filaments experience deceleration and acceleration respectively, leading to the difference in their energy spectrum~\cite{huynh2016}.  Fig.\ref{filaments} (a-b) shows density of the $(e_b^-, e_b^+)$ beam species with initial Lorentz factor $\gamma_\textrm{0}=50$ and a peak beam density per species $n_\textrm{b0}=\textrm{0.1}n_\textrm{0}$ after propagating a distance of 1500 $c/\omega_\textrm{p}$ (8 cm for an ambient plasma with $n_\textrm{0}=10^{16}\textrm{cm}^{-3}$) in an initially magnetized ($B_0=\textrm{7.5}\,$T) electron-proton plasma. One can see the filamentation of the pair-beam due to the WI/CFI in panels (a) and (b) and associated magnetic field energy in panel(c). Fig. \ref{filaments}(d) shows the difference in average kinetic energies between the beam $e_b^+$ and $e_b^-$ which is $\langle \gamma\rangle_{e_b^+}-\langle \gamma\rangle_{e_b^-}\approx 10\, (\sim 5\textrm{MeV})$. This energy difference is predominantly caused by the beam energy loss in generating the magnetic field, but also due to return current neutralization dynamics. Fig.\ref{enerdissip} shows that for density ratio $n_{b0}/n_0=0.1$, the $e_b^-$ looses energy while $e_b^+$ gains energy as explained before. However, at density ratio $n_{b0}/n_0=1$, both species loose energy to generate the magnetic field due to WI/CFI, though the $e_b^+$ population still has higher average energy than the $e_b^-$ population. Indeed, the beam to magnetic field energy conversion ratios in both cases read $2\%$ for $n_{b0}/n_0=0.1$ and $4.7\%$ for  $n_{b0}/n_0=1$. This suggests that $e_b^-$ population dominantly contributes to magnetic field generation. This can be simply understood by observing the fact that both the $e_b^+$ and return plasma currents are in the same direction and hence they are not as unstable as the oppositely moving $e_b^-$  and return plasma currents.

\begin{figure}
\centering
\includegraphics[width=0.45\textwidth, height=0.45\textwidth]{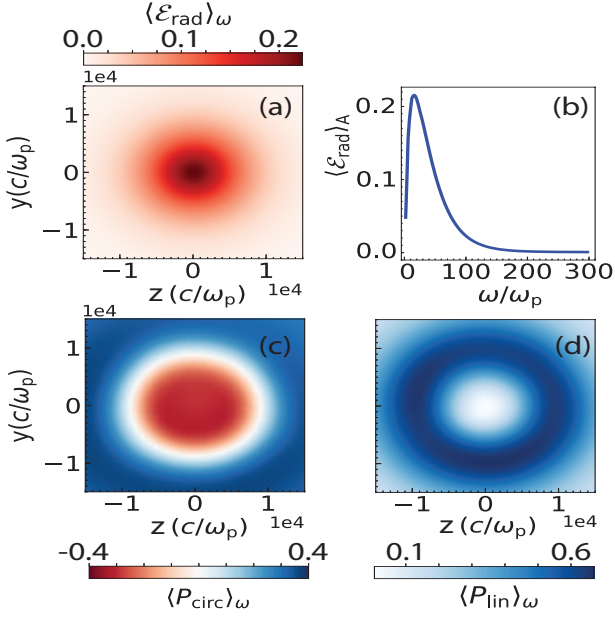}
\caption{Simulation results from CASPER illustrating the radiation characteristics from 1000 beam $e_\textrm{b}^-$ and $e_\textrm{b}^+$ taken together from time $t_i=750\omega_\textrm{p}^{-1}$ to $t_f=1500\omega_\textrm{p}^{-1}$ for $n_{b0}/n_0=0.1$ and $\gamma_\textrm{0}=50$. (a)-(b) Frequency and space averaged radiated energy respectively from the beam species. The $\langle \mathcal{E}_\textrm{rad}\rangle$ is normalized to $4\pi^2c/e^2$. Frequency averaged $\langle P_\textrm{circ}\rangle$ in panel (c), and $\langle P_\textrm{lin}\rangle$ in panel (d).}
\label{spectra}
\end{figure}
\begin{figure}
	\centering
	\noindent\includegraphics[width=0.46\textwidth,height=0.2\textwidth]{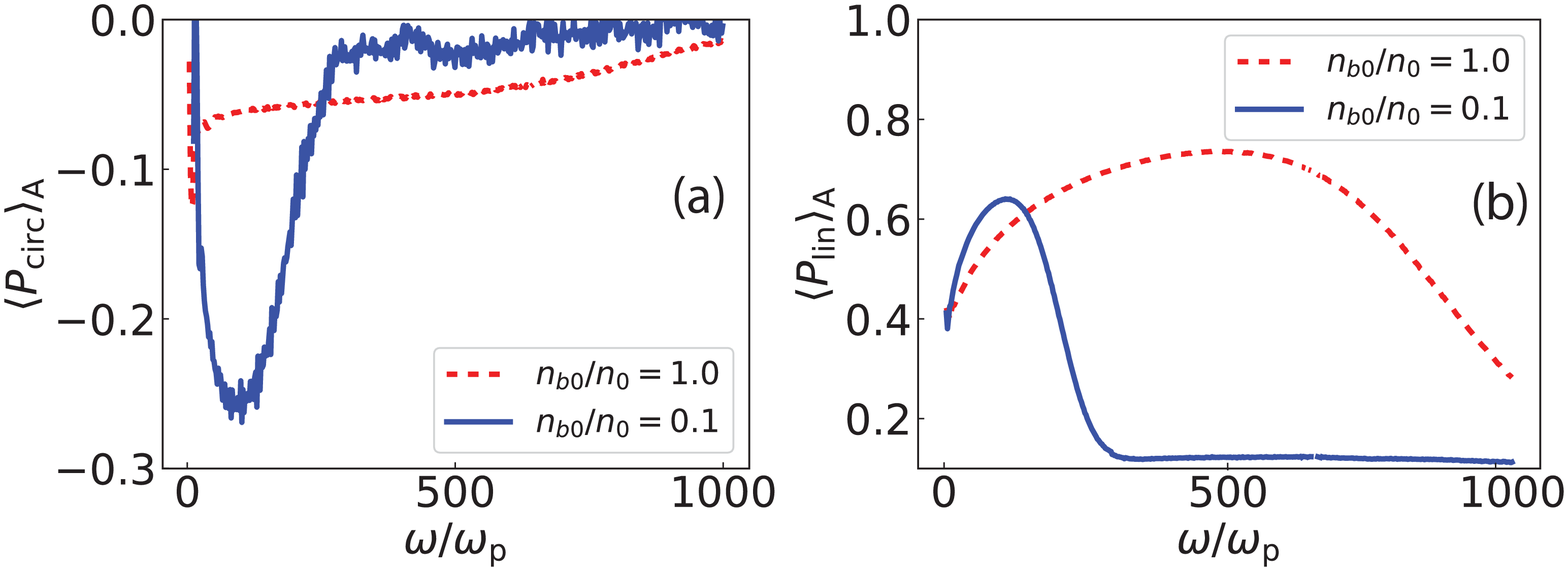}
	\caption{Frequency spectra of the polarized radiation for two cases of beam to plasma density ratios.}
\label{frequency}
\end{figure}

\paragraph{}
To analyze the spectrum and polarization of the radiation emitted from the beam particles, we extracted trajectories of 1000 beam $e_b^-$ and $e_b^+$ directly from the PIC simulations. The particles were randomly selected from those lying initially in the $xy$ plane at $z = \textrm{16}\,c/\omega_\textrm{p}$~\footnote{The average kinetic energy of the selected particles were comparable to that of the entire respective beam species, (see Supplementary material)}. We develop a new post-processing code CASPER~\cite{casper},  to compute the radiation spectrum and its polarization employing the method of Fourier transform of the radiated electric field and the Stokes parameters~\cite{sironi2009, hededal2005, Chen2013, martins2009}. A two dimensional virtual detector of size $[30000\times 30000] (c/\omega_{p})^2$ is kept in the $yz$ plane at a distance $x=10^5 (c/\omega_p)$, having $50\times 50$ grid points. The frequency range $\omega = (10^0-10^3)\omega_{p}$  (large enough for the spectrum) is analyzed with a resolution of $\Delta\omega=1$. To capture the influence of the filamentation, filaments merging and saturation of the WI/CFI magnetic fields, the radiation was calculated from the time $t_{i}=750\, \omega_{p}^{-1}$ to $t_{f}=1500\, \omega_{p}^{-1}$.  Defining $\boldsymbol{\epsilon}_1$ and $\boldsymbol{\epsilon}_2$ as the unit vectors perpendicular to the direction of observation and $\textbf{E}$ as the resultant radiated electric field vector from all the charged particles, the Stokes parameters are given as: $s_0 = |\boldsymbol{\epsilon}_1\cdot\textbf{E}|^2+|\boldsymbol{\epsilon}_2\cdot\textbf{E}|^2$; $s_1 = |\boldsymbol{\epsilon}_1\cdot\textbf{E}|^2-|\boldsymbol{\epsilon}_2\cdot\textbf{E}|^2$; $s_2 = 2\textrm{Re}[(\boldsymbol{\epsilon}_1\cdot\textbf{E})^*(\boldsymbol{\epsilon}_2\cdot\textbf{E})]$; $s_3 = 2\textrm{Im}[(\boldsymbol{\epsilon}_1\cdot\textbf{E})^*(\boldsymbol{\epsilon}_2\cdot\textbf{E})]$~\cite{Jackson:1998aa,born1970principles}. The total linear and circular polarizations read as $\langle P_\textrm{lin}\rangle=\langle\sqrt{s_1^2+s_2^2}/s_0\rangle$, and $\langle P_\textrm{circ}\rangle=\langle s_3/s_0\rangle$ respectively, where $\langle s_i/s_0\rangle = \iint s_i d\textrm{A}d\omega/\iint s_0 d\textrm{A}d\omega$ with $\textrm{A}$ as the area of the detector and $i=(1,2,3)$ \textcolor{black}{\footnote{It may be noted that these Stokes parameters are defined for a monochromatic wave and their applicability to an ensemble of plasma particles considered here is facilitated by the random phase approximation; see \cite{Rybicki:1985aa}.}}. Fig.\ref{spectra} (a-b) show the frequency and space averaged radiated energy from Fig.\ref{filaments}. The spectrum confirms the emission to be a synchrotron process. Since the $e_b^+$ and $e_b^-$ population have an energy difference and for a charged particle with Lorentz factor $\gamma$, the total energy radiated scales as $\mathcal{E}_\textrm{rad}\propto \gamma^4$~\cite{Jackson:1998aa}, the CP radiation flux (caused by $\mathbf{B_0}$) due to beam $e_b^+$ will exceed those from beam $e_b^-$. This can also be confirmed by comparing Figs.\ref{spectra} (c) and (d) where regions of high degree of circular polarization show small amount of linear polarization. The frequency averaged degree of circular and linear polarizations  with peak values of $\langle P_\textrm{circ}\rangle_\omega \approx 0.4 (40\%)$ and $\langle P_\textrm{lin}\rangle_\omega \approx 0.6 (60\%)$ are shown in Figs.\ref{spectra}(c) and (d), respectively. The total fractional linear and circularly polarized radiation fluxes (integrated over the whole area in Figs.\ref{spectra} (c) and (d)) are $\langle P_\textrm{circ}\rangle\approx 0.18 (18\%)$ and $\langle P_\textrm{lin}\rangle\approx 0.49 (49\%)$, respectively. In all cases, polarized light is found to be $\sim (67-69) \%$, making the fraction of unpolarized light to be $\sim (33-31) \%$.  Fig.\ref{frequency} shows the spectrum of the $\langle P_\textrm{circ}\rangle$ and $\langle P_\textrm{lin}\rangle$ peaks at $\omega \approx 100 \omega_\textrm{p}$ for $n_\textrm{b0}=0.1n_0$, which corresponds to a wavelength of $\sim\textrm{3}\,\mu\textrm{m}$ (far infra-red) for $n_0=10^{16}\textrm{cm}^{-3}$. This frequency emission is promising to explain the observation of optical circular polarization in GRBs. The spectrum broadens for higher $n_\textrm{b0}$ due to increased radiated energy, arising due to strong WI/CFI magnetic field.

\begin{figure}
	\centering
	\noindent\includegraphics[width=0.47\textwidth,height=0.38\textwidth]{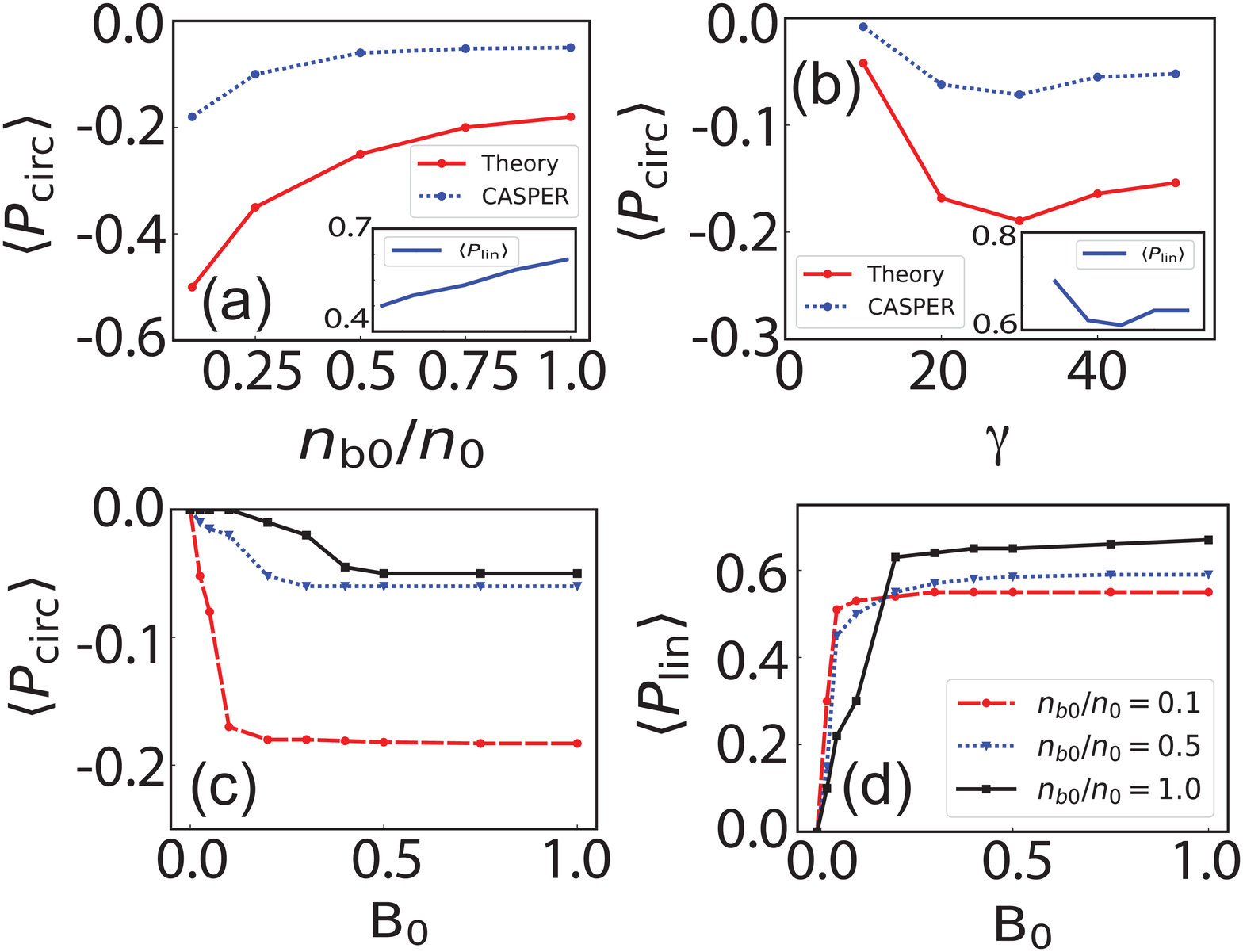}
	\caption{Scalings of $\langle P_\textrm{circ}\rangle$ from Eq.\eqref{pcscale} and PIC simulations. Scaling of the $\langle P_\textrm{circ}\rangle$ and $\langle P_\textrm{lin}\rangle$ (inset) (a) at  $\gamma_\textrm{0} = 50$, (b) at initial $n_\textrm{b0}/n_0 = 1$. For (a) and (b) $B_0=1$. Panels (c)-(d) show the scalings of $\langle P_\textrm{circ}\rangle$ and $\langle P_\textrm{lin}\rangle$ with $B_0$ at $n_\textrm{b0}=[0.1, 0.5, 1.0]$ and initial $\gamma_\textrm{0}=50$.}
\label{scalings}	
\end{figure}

In order to check the validity of our simulation results, we use the well-known synchrotron radiation formalism~\cite{Jackson:1998aa} and calculated the degree of polarization analytically. The Fourier spectrum of the radiation field is, $\textbf{E}(\textbf{r},\omega)=2\,e\, \omega\big [(1+\gamma^2\theta^2) \, K_{2/3}(\textrm{x})\hat{\textbf{e}}_1 / (\sqrt{3}\gamma^2\omega_0) \pm i\theta {\sqrt{1+\gamma^2\theta^2}}\, K_{1/3}(\textrm{x})\hat{\textbf{e}}_2/ (\sqrt{3}\gamma\omega_0)\big] \textrm{e}^{i\omega r}/ ( c\,r)$, where $e \equiv (e_b^-,e_b^+)$, $+$ is for $e_b^+$ and $-$ for $e_b^-$, $x=\omega/(3\gamma^3\omega_0)(1+\gamma^2\theta^2)^{3/2}$, $\omega_0$ is the cyclotron frequency, $r$ the distance from the charge to the observation point, and $K_{1/3}(x)$ and $K_{2/3}(x)$ are the modified Bessel functions. Thus, the radiation will be circularly polarized for $\theta\neq 0$. Numerically integrating the Stokes parameters $s_3$ and $s_0$ for the radiated field given before over the entire solid angle and the frequency range, we obtain the total circularly polarized radiation flux, $P_\textrm{cflux} = 7.2(e/c)^2\gamma^4$ and the total radiated energy $\mathcal{E}_\textrm{rad}=8.556(e/c)^2\gamma^4$ for a single particle. For the pair beam with distribution function from PIC simulations, $f(\gamma)_\alpha$, where $\alpha=(e_b^-, e_b^+)$, the resultant degree of circular polarization  
$\langle P_\textrm{circ}\rangle = (\langle P_\mathrm{cflux}\rangle_{e_b^-}-\langle P_\mathrm{cflux}\rangle_{e_b^+})/(\langle \mathcal{E}_\mathrm{rad}\rangle_{e_b^-}+\langle \mathcal{E}_\mathrm{rad}\rangle_{e_b^+})$, where $\langle P_\mathrm{cflux}\rangle_{\alpha}=\int P_\textrm{cflux}f(\gamma)_{\alpha}d\gamma/\int f(\gamma)_{\alpha}d\gamma$, yields
\begin{equation}\label{pcscale}
\langle P_\mathrm{circ}^{\textrm{th}}\rangle = 0.8415\frac{\langle \gamma^4\rangle_{e_b^-}-\langle \gamma^4\rangle_{e_b^+}}{\langle \gamma^4\rangle_{e_b^-}+\langle \gamma^4\rangle_{e_b^+}}.
\end{equation}
Eq.\eqref{pcscale} indicates that a finite circular polarization arises due to the difference in average kinetic energies between the beam species. For the purpose of comparison of PIC simulation results and Eq.\eqref{pcscale}, we performed two sets of PIC simulations: First, with $n_\textrm{b0}/n_0=[0.1, 0.5, 1.0]$, keeping the beam initial $\gamma_\textrm{b0}=50$ and then with $\gamma_\textrm{b0} =[10, 20, 30, 40, 50]$, keeping the $n_\textrm{b0}/n_0=1.0$. The $\langle \gamma^4\rangle_\alpha$ of the beam species was calculated for the duration $(750-1500)\omega_\textrm{p}^{-1}$ using the distribution obtained from the simulation data. The $\langle P_\textrm{circ}\rangle$ followed the scaling of Eq.\ref{pcscale} as shown in Fig.\ref{scalings} (a-b). The $\langle P_\textrm{circ}\rangle$ decreases while the $\langle P_\textrm{lin}\rangle$ (insets) increases with $n_\textrm{b0}/n_0$. Since asymmetric energy dissipation in $e^+_b$ and $e^-_b$ species is larger at low density (as seen in Fig.\ref{enerdissip}), the CP flux is also higher at lower densities. At higher density the energy difference between beam species is lower, yielding strong linear polarization as expected from synchrotron emission \cite{Rybicki:1985aa,Medvedev:1999aa}.    \textcolor{black}{Contrastingly, $\langle P_\textrm{circ}\rangle$ increases ($\langle P_\textrm{lin}\rangle$ decreases) with $\gamma$ (peaks around $\gamma_0 =30$ ) and decreases afterwards as shown in Fig. \ref{scalings}(b) for density ratio $n_{b0}/n_0 =1$. The decrease can be associated with the effect of plasma hosing and modulation instabilities in plasma wakefield regime which may further reduce and saturate the $\langle P_\mathrm{circ}\rangle$ for high beam densities.} Panels (c) and (d) show the {dependence} on the external magnetic field $B_0$ for different initial $n_\textrm{b0}$. Both components of polarizations increase with $B_0$ and saturate when the  WI/CFI generated magnetic field equals the external magnetic field. The WI/CFI fields add a degree of randomness to the motion of charged particles reducing $\langle P_\mathrm{circ}\rangle$ and $\langle P_\mathrm{lin}\rangle$. For high $B_0$, the motion of the beam particles is dominated by $B_0$ and $\langle P_\mathrm{circ}\rangle$ arises only due to the energy difference between the beam $e_b^-$ and $e_b^+$ as seen in Fig.\ref{scalings} (c-d). In case of an $e_b^-,e_b^+$ ambient plasma, both the species of the ambient plasma contribute equally to neutralize the current filaments of the beam species. As a result, there is no asymmetry in the kinetic energies between the beam species. Then $\langle P_\textrm{circ}\rangle = 0$ for such a case and is independent of $n_\textrm{b0}/n_0$, as the $\langle P_\textrm{circ}\rangle$ due to the beam $e_b^-$ is canceled by that due to the beam $e_b^+$; see also Ref.~\cite{Sinha:2015aa} where generation of circular polarization in a different physical configuration is attributed to the topological changes in pitch-angle distribution of electrons instead of the asymmetric energy dissipation  discussed here. 

In the PIC simulations, we have not included the electron-positron annihilation and radiation reaction effects. The annihilation rate in a plasma, in the ultra-relativistic limit, can be written as $\dot{n}_{+}=(2\pi c\, n_{+}n_{-} r_e^2/\gamma_0)\left[\mathrm{ln}\,2 \gamma_0 -1\right]$ sec$^{-1}$, where $n_{+}, n_{-}$ are the electron and positron densities respectively, $r_e$ is the classical electron radius, and $\gamma_0$ is the Lorentz factor of the pair beam~\cite{Svensson:1982aa}. For PIC simulation parameters, $n_{+}=n_{-}=10^{15}$cm$^{-3}$, and $\gamma_0=50$,  it yields $\dot{n}_{+}\sim 1$ fs$^{-1}$, justifying the assumption of ignoring the pair annihilation in our PIC simulations. Also, the magnetic field due to the WI doesn't grow to large values to necessitate the inclusion of RR force in our PIC simulations~\footnote{We carried out PIC simulations by including the Landau-Lifshitz radiation reaction force and found results to be unchanged in agreement with the results of Ref.~\cite{DAngelo:2015aa}}. \textcolor{black} {Also we have ignored that plasma collisions due to lower plasma density employed in PIC simulations}.

In conclusion, we have shown that a pair beam ($e_b^-, e_b^+$) propagating through a magnetized electron-proton plasma emits radiation by the synchrotron mechanism. The asymmetry in the kinetic energies of the beam species arising due to the energy dissipation into generating the transverse magnetic field do not allow the left and right circularly polarized radiation fluxes from the beam species to cancel, resulting in a finite circularly polarized radiation flux. The degree of linear and circular polarization depend on the initial $n_{b0}$ and $\gamma_0$ of the fireball, the initial magnetization and the composition of the ambient plasma. These results can be readily tested in laboratory astrophysics experiments. In addition, the origin of circular polarization due to asymmetric energy dissipation between beam $e_b^-$ and $e_b^+$ species opens the scope to explain the recently observed  circular polarization in the optical frequency range from GRBs~\cite{wiersema2014}.

\begin{acknowledgments}
The authors would like to thank Dr. K. Hatsagortsyan for useful discussions. U. Sinha would like to thank the SMILEI development team for technical support.
\end{acknowledgments}


\begin{thebibliography}{40}
\bibitem [{cas()}]{casper}%
  \BibitemOpen
  \href@noop {} {}\bibinfo {note} {It is a parallel C code that is used to
  compute radiation spectrum and its polarization on a virtual detector by
  postprocessing the trajectories of particles obtained from PIC simulations.
  We term it as CASPER which stands for Coherence And Spectral Postprocessor
  for Emitted Radiation. It can be directly used to postprocess for PIC codes
  SMILEI and EPOCH. It has a separate trajectory restructuring routine that can
  be modified to accomodate trajectory data files from other PIC codes
  too.}\BibitemShut {Stop}%
\end{thebibliography}
%

\end{document}